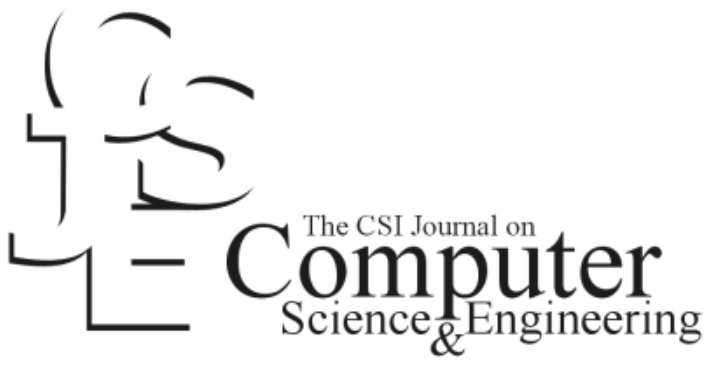

# Multi-Sensor Fuzzy Data Fusion Using Sensors with Different Characteristics

Mohammad Amin Ahmad Akhoundi [1], Ehsan Valavi[1,2]

[1]University of Tehran Alumni, [2] IPM

## Abstract

This paper proposes a new approach to multi-sensor data fusion. It suggests that aggregation of data from multiple sensors can be done more efficiently when we consider information about sensors' different characteristics. Similar to most research on effective sensors' characteristics, especially in control systems, our focus is on sensors' accuracy and frequency response. A rule-based fuzzy system is presented for fusion of raw data obtained from the sensors that have complement characteristics in accuracy and bandwidth. Furthermore, a fuzzy predictor system is suggested aiming for extreme accuracy which is a common need in highly sensitive applications. Advantages of our proposed sensor fusion system are shown by simulation of a control system utilizing the fusion system for output estimation.

## 1. Introduction

Measurement is a significant requirement for today's industrial world. Applications such as control, safety and monitoring systems, which are the inseparable parts of any industry, are extensively using measurement systems. In addition, on the product development side, today's innovative solutions involve multiple modules and components and due to this complexity, the final product requires highly accurate measurement systems to guarantee performance. For example, in a typical passenger airplane there are thousands of sensors and in a typical factory there are hundreds of them. The more advances are made in industries, the more demand is observed for more accurate measurements. This soaring use of measurement solutions lead to a great deal of research in increasing their performance.

Although many parameters are in play in restricting measurement system's performance, the main contributor in limiting their potentials is lack of sensors that meet certain requirements. The restrictions originate from the physical features of a sensor and can be witnessed in fabrication problems or where replacing old sensors are not economically justifiable. Despite of these hardware limitations, the performance can still be improved using software solutions. For example, in the fabrication process there are always room for error and this manufacturing error causes variation in performance characteristics of produced sensors. However, using multiple sensors in parallel can enhance the quality of the measurement by reducing the measurement error's effect. As another example, in oceanography or other earth science disciplines, new and more accurate sensors are always needed. But since replacing old sensors are not economically justifiable, new sensors work in parallel with the old ones. The information that are measured by those old sensors, although not as accurate as the gathered information by new ones, are still useful and can be used to enhance the quality of new sensor's measurements.

Sensor-fusion is a software approach for improving reliability of information obtained from a sensory system. It uses aggregation of multiple sensors' information to produce enhanced quality measurements. Aggregated information (Also referred as optimal or maximum information [1]) conveys valuable and reliable data that cannot be explicitly found in primary sensor measurements. By aggregation of information, sensor-fusion reduces the effect of each sensor's errors and erasures which ultimately leads to a better and more accurate estimation of the measured variable.
Fusion methods can be categorized into three major clusters.

- Probabilistic methods such as Bayesian analysis of sensor values, Evidence Theory, Robust Statistics, and Recursive Operators.
- Least square-based estimation methods such as Kalman Filtering, Optimal Theory, Regularization, and Uncertainty Ellipsoids.
- Intelligent aggregation methods such as Neural Networks, Genetics Algorithms, and Fuzzy Logic.

In most research in this area, all sensors have been treated similarly, i.e. no well-defined differences in sensor's performance characteristic are considered. In this research we propose a fuzzy intelligent method for sensor fusion which is mainly based on information about sensor's characteristics. We specially concentrate on accuracy and bandwidth of an ordinary sensor as the parameters having the most effects on the performance of a sensory system. The reason for this consideration is that many applications require sensors with high accuracy and enough sensitivity in response to rapid and high-frequency changes in the measured variable. In some cases, such sensors are not easily available or using them are indeed not economically justifiable. Our solution to this need is based on aggregation of information from sensors with complement characteristics in accuracy and bandwidth, in order to have an acceptable estimation of the measured variable that meets our requirements. Despite many other methods, our fusion algorithm does not need system's model which the measured variable belongs to. It only depends on some general information about accuracy and bandwidth of the sensors. Since the problem of inaccuracy and slowness of sensors appears mostly in industrial control systems, our focus in this paper is on control applications.

The rest of this paper is organized as follows. A summarized literature review is presented in Section 2. Problem definition is presented in Section 3 where we also introduce our method and explain its two components in detail. Simulation of the method in a control system benchmark and discussion about its results is in Section 4. At the end, Section 5 summarizes this paper with conclusion and discussion.

# 2. Related Work

A complete survey of information fusion techniques for reliable data fusion can be found at [2], however, sensor fusion problems, applications, and future directions are completely addressed at [3,4,5,6,7]. As it can be grasping from contents of mentioned papers, industrial need triggered current research stream on more precise, yet feasible, algorithms for sensor fusion. Consequently, we can see many application-oriented uses of sensor fusion in the literature. A case in point, in [8] they mainly concentrate on reducing redundancy and noise and attempted to improve failure tolerability of information generated by sensors of gas turbine power plants. More practical and industrial application of sensor fusion can be found at [9, 10, 11, 12].

Those precise mathematical algorithms leveraged probabilistic models, evolutionary methods, and intelligent decision-making means as discussed and categorized previously. An optimal linear fusion framework was proposed in [13] for addressing and solving measurement systems' problems. In [14], authors consider two-sensor signal enhancement problem in a noisy environment. The proposed solution is based on expectation maximization algorithm for jointly estimating the main signal, the coupling system and the unknown signal and noise parameters. [15] presents a systematic scheme for generating optimum fusion rules, which reduces computation tremendously compare to ordinary exhaustive search. In [16], two novel neural data fusion algorithms based on a linearly constrained least square (LCLS) method are proposed.

Early attempts for recruiting fuzzy rules in multisensory data fusion can be seen at [17]. Inspired by their opinions, other researchers use fuzzy rules for overcoming mathematical approaches' shortcomings in proposing practical solution. For instance, a fuzzy-based multi-sensor data fusion classifier is developed and applied to land cover classification using ERS-1/JERS-1 SAR composites in [18]. In addition to using fuzzy rules, research such as [19, 20] have tried to introduce new fuzzy logic operators in order to utilize intuitive knowledge about a system for sensor-fusion. In a practical problem, to correct slow sensor drift faults, [21] presented a hybrid method using fuzzy logic and genetic algorithm. In order to exploit advantages of both fuzzy-logic as an outstanding intelligent method, and Kalman filter as an efficient fusion method, [22] suggests a hybrid Kalman filter-fuzzy logic adaptive multi-sensor data fusion architectures.

In most of these researches, no clear differences between sensors is considered, but sensors have certain known characteristics describing their behaviour. Naturally, using information about these characteristics can lead to better design of sensor fusion algorithms. The results may be more robust, if instead of using algorithms depending on system's model, we take advantages of methods considering sensor characteristics only.

# 3. Proposed Methods

## 3.1. Problem Statement

The general goal of the system is to measure physical variable of interest as accurate as possible. We use a wideband sensor (S1) that is not accurate enough to be used alone and a more accurate sensor with lower bandwidth (S2). These sensors have been utilized for a control system. The role of proposed method is to aggregate information obtained from these two sensors to best estimate the physical variable as a feedback for control purposes.

## 3.2. The General Structure of Sensor-Fusion Method

The system uses the general structure described in Fig. 1. The Fuzzy Aggregator is the main part of this system. In a real-time process, it uses sampled data from $S_1$ and $S_2$ to produce an estimate of the measurements. Fuzzy Predictor (FP) may also be used in the system for better estimation. The Fuzzy Predictor uses prior samples of estimated values to predict the upcoming value. Despite of FP's benefits in applications needing an extreme accuracy, it is not an essential component of the structure and may be omitted from the design depending on application's requirement. It is because of the fact that its complexity in calculations leads to problems in real-time processing which makes it a problem rather than a solution in some applications. Therefore, Fuzzy Aggregator can be used without Fuzzy Predictor.

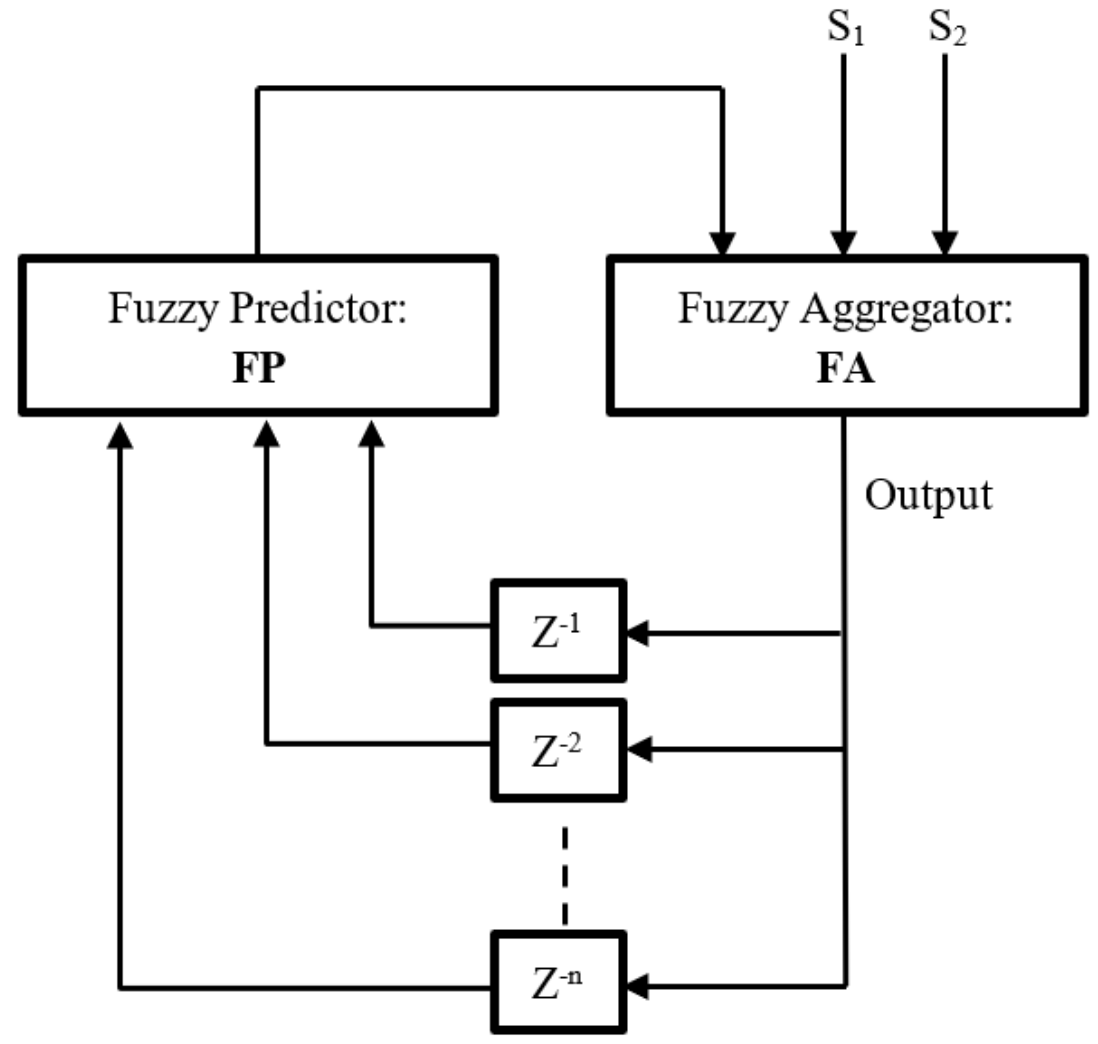

Fig. 1. The general structure of our sensor fusion method

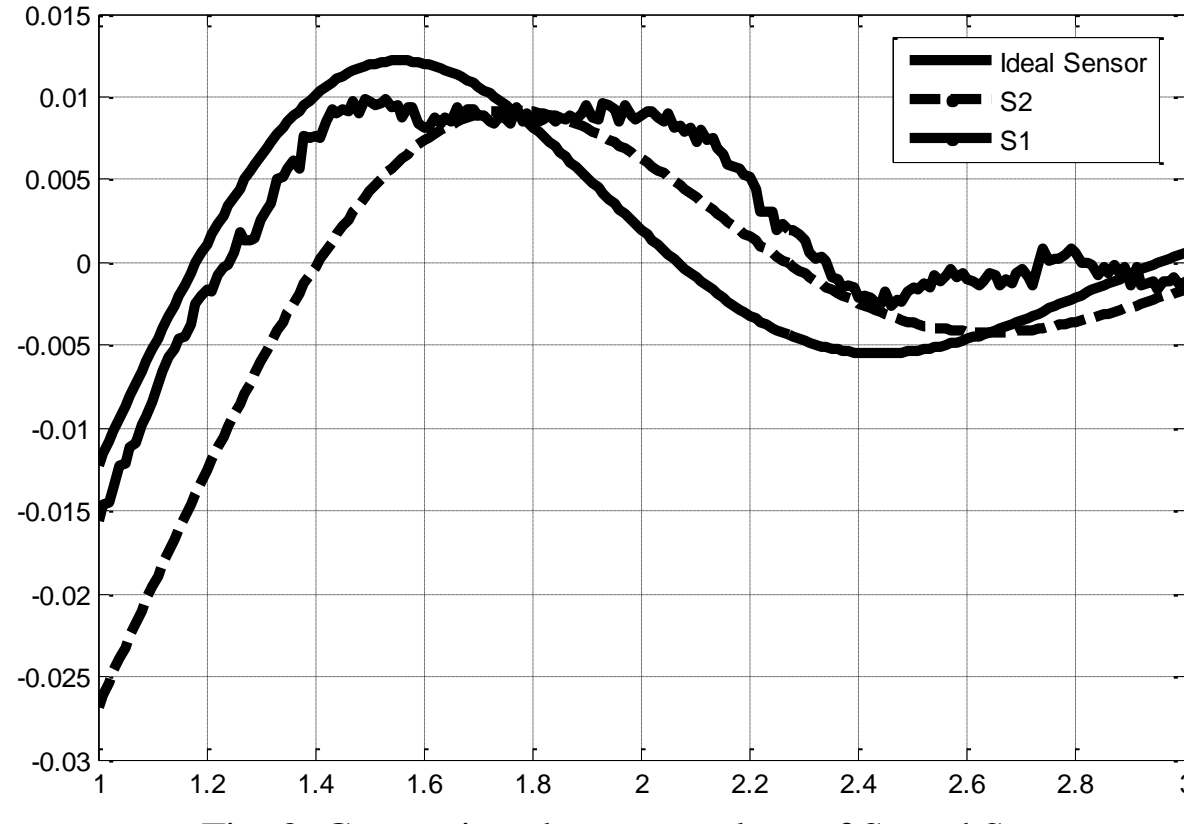

Fig. 2. Comparison between values of $S_1$ and $S_2$

## 3.2. Fuzzy Aggregation

Without considering Fuzzy Aggregator's predicted input, we can deal with only $S_1$ and $S_2$. A weighted average of $S_1$ and $S_2$ can provide an acceptable estimation of measured variable, if the weight of each sensor is determined appropriately.

$$f(S_1, S_2) = \frac{w_1 S_1 + w_2 S_2}{w_1 + w_2} \tag{1}$$

Fuzzy aggregator (FA) determines weight of each sensor in (1) through a rule-based fuzzy system. Assuming the sensors' weights are normalized, the fuzzy system is required to calculate only the weight of $S_1$ as the output.

Differential of $S_2$ measurement and difference between $S_1$ and $S_2$ values are selected as the two inputs of the fuzzy system. Differential of $S_2$ through boost of high frequency components, regains some of the signal's information loss by the low-bandwidth sensor. Because of its low noise, it also provides an appropriate value for judging the rate of changes in measured variable. Due to its inaccuracy, S1 can give misleading information about the signal changes, but its difference with measured of S2 can give some information about the reliability of S1 in situations with low changes in value of S2.

For determining the weights of each sensor, we must use system's inputs appropriately. S1 must have a greater weight, when the changes in the measured variable are rapid, and these changes cannot appear in S2 due to its slow response. In addition, when the variable does not have rapid alternations and it changes in a smooth way, the weight of S2 should be greater to prevent the uncertainty of $S_1$ from affecting on the estimation. On the other hand, if the slope of changes in the measured variable is slow, and the difference between the values of S1 and S2 is large, it is because of high inaccuracy of S1, and therefore its weight should be highly reduced. In case of high changes in the measured variable and also high difference between measured values, it can be concluded that mentioned difference has happened due to slow response of S1; therefore, weight of S1 should be very large. These inferences can be expressed as fuzzy rules:

If $abs\ (S_1 - S_2)$ is small and $abs\ (dS_2/dt)$ is small,
Then: W1 should be small.

If $abs\ (S_1 - S_2)$ is small and $abs\ (dS_2/dt)$ is large,
Then: W1 should be large.

If $abs\ (S_1 - S_2)$ is large and $abs\ (dS_2/dt)$ is small,
Then: W1 should be very small.

If $abs\ (S_1 - S_2)$ is large and $abs\ (dS_2/dt)$ is large,
Then: W1 should be very large.

In addition to W1, the fuzzy system should also have another output. The weighted average will be an appropriate estimation, only if the real signal lies between the two sensors' values. There are situations in which error in value of S1 leads both values associated with S1 and S2 to be placed in one side of the real signal. Such a situation is more "probable" when rate of measured variable is high, but the difference of values of S1 and S2 is still low. We need a new variable as another output of Fuzzy Aggregator to compensate this error. The output variable called "drift" is an estimation of the amount of error and is added or subtracted from the final estimation according to the slope of changes in the measured variable. The structure which is required for this operation can be described in Fig. 3. The fuzzy rule that we can obtain from these linguistic analyses can be expressed as below:

Drift should be large only when:
If $abs\ (S_1 - S_2)$ is small and $abs\ (dS_2/dt)$ is large.

Table 1 presents obtained fuzzy rules. As the "common sense" fuzzy rules are obtained, appropriate membership functions for linguistic variables used in the fuzzy rules should be defined. Fig. 4 shows selected membership functions. The domain of function definitions depends on the expected range of changes in the measured variable.

Table 1. Fuzzy Rules

| Inputs | | Outputs | |
|---|---|---|---|
| Abs($S_1$-$S_2$) | Abs($dS_2$/dt) | $W_1$ | Drift |
| S | S | S | S |
| S | G | G | G |
| G | S | SS | S |
| G | G | GG | S |

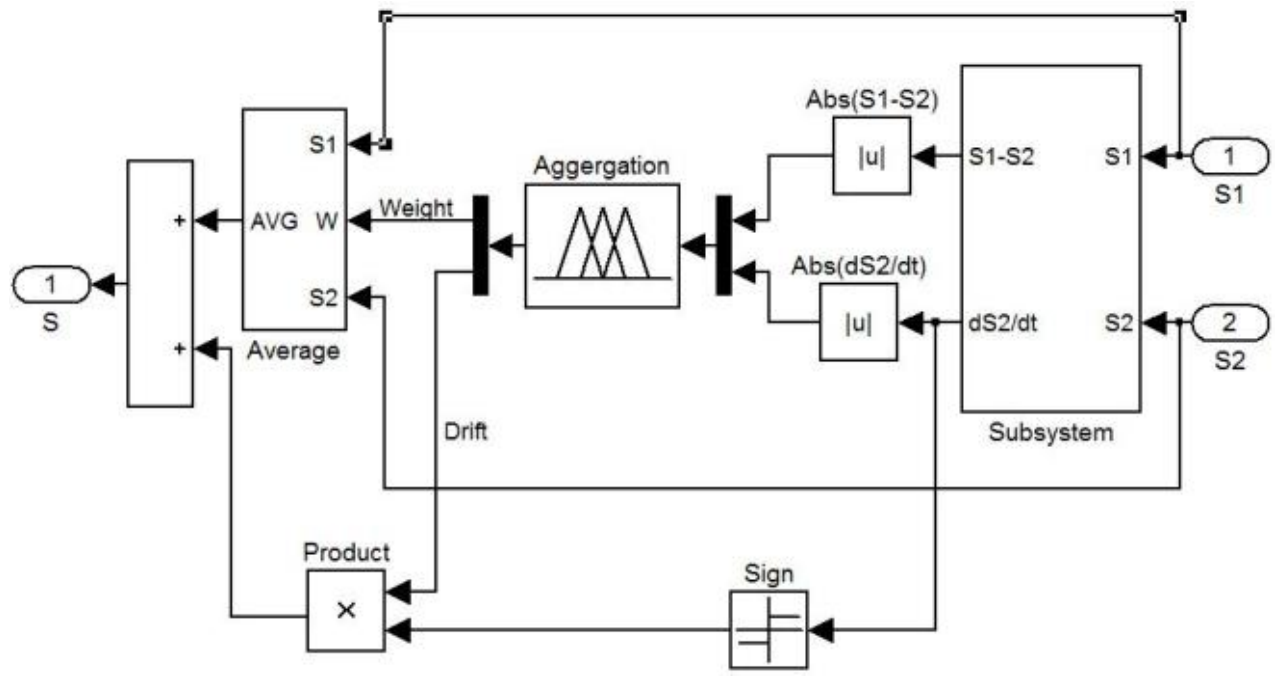

Fig. 3. General Structure of the Rule-Based Fuzzy System

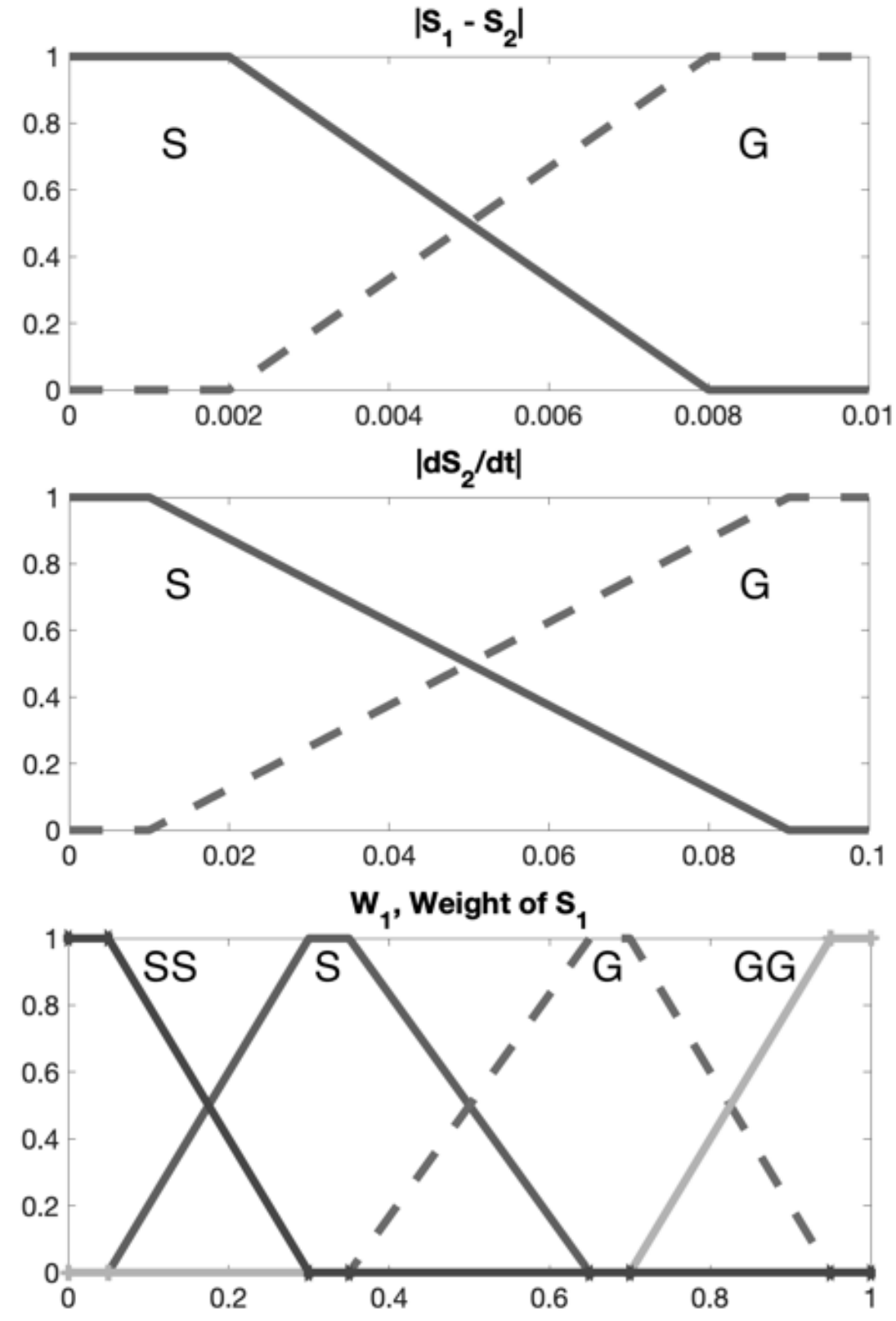

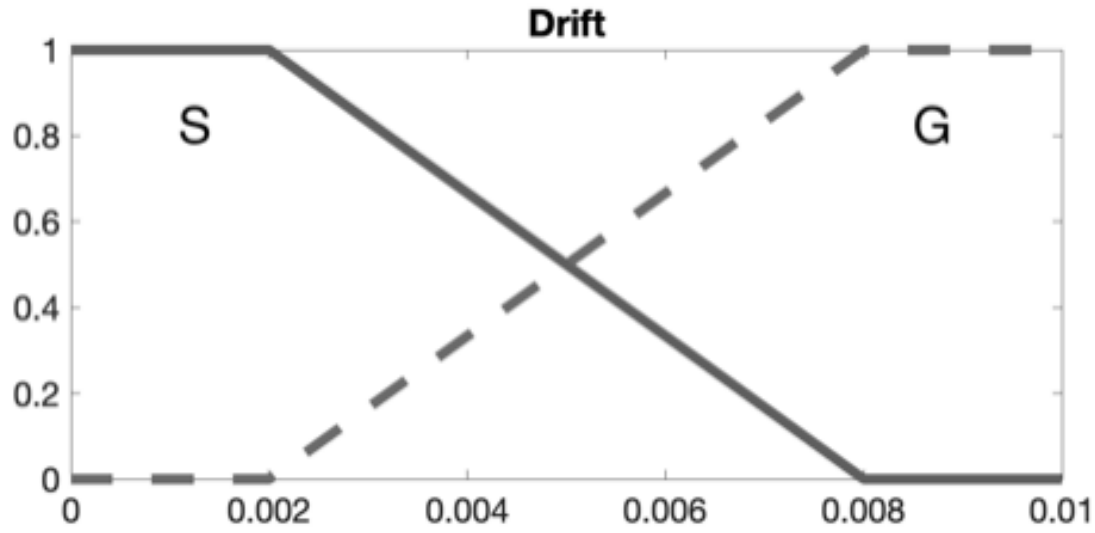

Fig. 4. Membership functions for inputs and outputs of the system

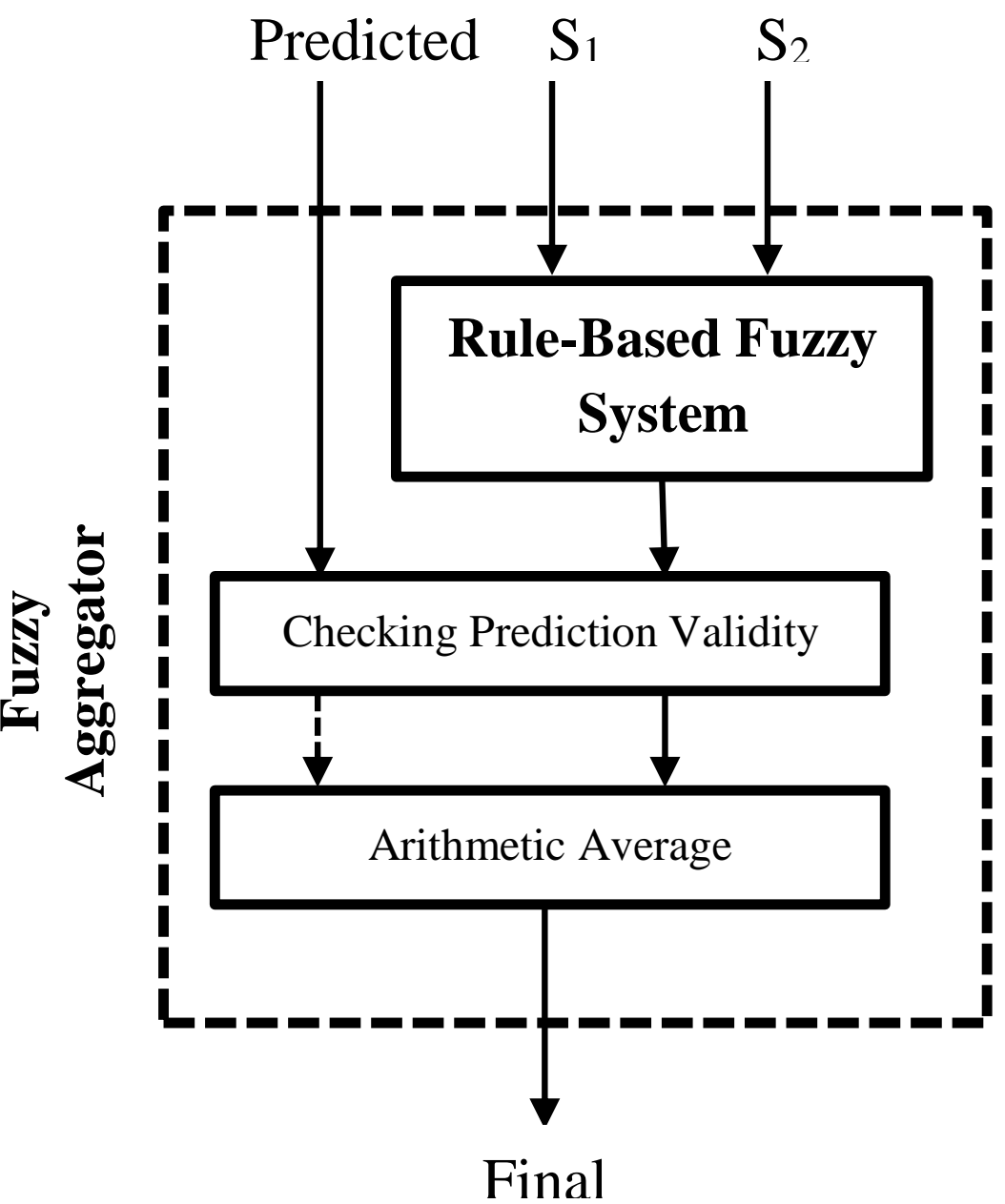

Fig. 5. General Scheme of Fuzzy Aggregator

What is the role of the predicted input? It is an input for improving the fuzzy fusion system's performance. The way by which its value should be obtained is explained in next section. The key point here is that it is just a prediction and hence, it is not always reliable. In environments with strongly changing and repeated stochastic disturbances, the predictor's performance deteriorates, and it is not going to be very useful. However, in systems needing extreme accuracy (even at the expense of complexity of calculations) and also with smoothly changing disturbances, the fuzzy predictor can be beneficial. In the time ranges of happening disturbances, the predictor might be wrong. We can compare the prediction with the output of fuzzy system, as depicted in Fig. 5, to check the validity of prediction. The difference toleration level between the predicted value and the output of the rule-base fuzzy system should be defined according to the application types. If the result of the check exceeds the level, it should be ignored; otherwise the arithmetic average of its value and the output of rule-based fuzzy system should be considered as the final output of the Fuzzy Aggregator system. Fig. 5 shows the exact and complete structure of the Fuzzy Aggregator system.

## 3.3. Fuzzy Aggregation

### 3.3.1. System's Overview

From conceptual perspective, designing fuzzy systems based on input-output data can be divided into two categories of methods. In the first category, the fuzzy system is designed based on the fuzzy rules obtained from the input-output data. Design of the Fuzzy Aggregator (FA) has been done from this perspective. But the Fuzzy Predictor uses the second method. It suggests choosing an appropriate structure for the system (naturally including some parameters), and then optimizing parameters using an appropriate training algorithm.

The Fuzzy Predictor system is an intelligent system with $n$ inputs, which are $n$ previous consecutive samples of the Fuzzy Aggregator's outputs. The system's structure can be expressed in (2). In following formula, $x_1, x_2 \dots x_n$ are "n" previous samples.

$$f\ (x_1, x_2, \dots, x_{n-1}) = x_n \tag{2}$$

In each sampling period, which is a training step, system's parameters are modified aiming to achieve closer estimation of $f\ (x_1, x_2, \dots, x_{n-1})$ to $x_n$ . Using this online training algorithm, we have an extrapolation that is $f\ (x_2, x_3, \dots, x_n) = x_{n+1}$, predicting the next sample of the measured variable. After passing enough training steps, the prediction can be reliable meaning that realized and predicted $x_{n+1}$ are sufficiently close to each other.

### 3.3.2. Gradient Descent Algorithm

Assume a differentiable function $l(x) \in \mathbb{R}^N$ where $x$ is a vector representing its inputs. Also assume that this function has $N$ unknown parameters, $p_1, p_2, \dots p_N$. The goal is to make $l(x)$ as small as possible at a certain point $x$, by choosing optimal values for the parameters. The gradient descent algorithm proposes an optimization method to achieve this goal. It suggests that function's descent direction to the optimal parameters can be characterized by the gradient of function $l(x)$. The algorithm hence starts with some initial values for the parameters $p_1, p_2, \dots p_N$, and at each step it moves in the opposite direction of the gradient. It continues this process till it reaches parameters where no meaningful decrease in $l(x)$ can be observed. Following equation shows the updating rule in each step.

$$p_i(q+1) = p_i(q) - \alpha \left.\frac{\partial l(x)}{\partial p_i}\right|_q \tag{3}$$

Where $i \in \{1, \dots N\}$. The symbol $q$ demonstrates the training step and $\alpha$ represents the step size. For our particular problem, for the sake of simplicity, we assume the stepsize to be a constant. This constant value should be therefore small enough to lead to convergence of algorithm.

### 3.3.3. Fuzzy Predictor Using Gradient Descent

By choosing different forms of system's membership functions, we can obtain different classes of the fuzzy systems. If we choose multiplying inference engine, singletone fuzzifier, center average deffuzifier, and Gaussian membership functions, the general form of such system can be written in the close-form expression (4).

$$f(x) = \frac{\sum_{l=1}^{M} \bar{y}^l \left[\prod_{i=1}^{n} \exp(-(\frac{x_i - \bar{x}_i^l}{\sigma_i^l}))\right]}{\sum_{l=1}^{M} \prod_{i=1}^{n} \exp(-(\frac{x_i - \bar{x}_i^l}{\sigma_i^l}))} \tag{4}$$

In which input x is a vector in $\mathbb{R}^n$ representing n inputs of the system. The output f is also signal's "upcoming sample" predictor and parameter M shows number of rules. Such structure has the capacity of showing intelligent behaviors due to fuzzy systems' intrinsic features. The system's unknown parameters are:

$\bar{y}^l\ (l = 1,2 \dots M)$
$\sigma_i^l\ (l = 1,2 \dots M, i = 1,2, \dots n)$
$\bar{x}_i^l (l = 1,2 \dots M, i = 1,2, \dots n)$

Considering following definitions, we can simplify above expression.

$$z^l = \prod_{i=1}^{n} \exp(-(\frac{x_i - \bar{x}_i^l}{\sigma_i^l})) \tag{5}$$

$$a = \sum_{l=1}^{M} \bar{y}^l z^l, \qquad b = \sum_{l=1}^{M} z^l \tag{6}$$

$$f = a/b \tag{7}$$

In each sampling period, system's goal is to minimize the error or loss defined in following way.

$$l = \frac{1}{2} \| f(x_1, x_2, \dots, x_{n-1}) - x_n \|_2^2 \tag{8}$$

According to gradient descent method in each training step we should modify the parameter in this general way:

$$p(q+1) = p(q) - \alpha \left.\frac{\partial l}{\partial x}\right|_q \tag{9}$$

Using (4-7) we can obtain recurrence expressions required to modify systems' parameters in each training step:

$$\bar{y}^l(q+1) = \bar{y}^l(q) - \alpha \frac{f-y}{b} z^l \tag{10}$$

$$\bar{x}_i^l(q+1) = \bar{x}_i^l(q) - \alpha \frac{f-y}{b} (\bar{y}^l(q) - f) z^l \frac{2\left(x_i^p - \bar{x}_i^l(q)\right)}{\bar{\sigma}_i^{l^2}(q)} \tag{11}$$

$$\bar{\sigma}_i^l(q+1) = \bar{\sigma}_i^l(q) - \alpha \frac{f-y}{b} (\bar{y}^l(q) - f) z^l \frac{2\left(x_i^p - \bar{x}_i^l(q)\right)^2}{\bar{\sigma}_i^{l^3}(q)} \tag{12}$$

Modification of parameters according to (10-12) continues until the amount of error (e) becomes smaller than a threshold. Then the predicted value as the output of the system will be calculated as (13).

$$f\ (x_2, x_3, \dots, x_n) = x_{n+1} \tag{13}$$

Next sampling period uses same procedure on new set of inputs.

# 4. Simulation and results

Accuracy and speed of measurement system highly affects performance of control systems. If the feedback provided for a control system is not accurate and rapid enough, the system will fail to regulate the output properly. Besides, choose of inappropriate sensors leads to system's oscillation and even its instability. Due to the significant role of measurement systems in control applications, our focus for assessing the fusion method is on response evaluation of the control system that uses it. As a benchmark we chose an inverted pendulum control system for evaluation of proposed measurement system.

Figures 6 through 18 show response of the control system and also outputs of the fuzzy system designed for sensor fusion. The system has an initial condition and is required to set the output at zero. A disturbance is applied to the system after 25 seconds. Using an ideal sensor causes the system's response to be similar to what is shown in Fig. 6. If we only use sensor S1 as the feedback, the system's response is similar to what is shown in Fig. 7. The inaccuracy of this sensor causes high deviation in the response from the zero line. Use of a slow sensor S2 can lead to an oscillatory response as depicted in Fig. 8. If we use an ordinary average to combine information from S1 and S2, the response would be similar to what is shown in Fig. 9. This illustrates that a simple aggregation method, in which weights of sensors are equal in all time, is not efficient and consequently, a more intelligent method is required to achieve an acceptable performance.

In previous sections, it was mentioned that the fuzzy system including a rule-based system (that is a part of fuzzy aggregator in presence of predicted input) and a fuzzy predictor can be used to produce an accurate estimate of the measured variable. It was then claimed that in many applications, the fuzzy aggregator can be the only component used for fusion. This component has two outputs: W1 which is weight of S1 and the Drift. The former is the main output of this component, and the latter is used to compensate probable error that may happen in the system. We discussed that system without the Drift output may have unacceptable performance. Fig. 10 provides an evidence to support this argument. In this figure, system's response with the fuzzy aggregator and without the Drift output is shown. Fig. 11 supports the argument that a complete Fuzzy Aggregator system (with both Drift and W1 outputs) can help to achieve an appropriate response. In this simulation (For Fig. 11) we didn't have the fuzzy predictor. Figures 12 and 13, show the changes in Fuzzy Aggregator's outputs. Fig. 14 compares the real signals and values of S1 and S2. It can be observed that sensor S1 has not only considerable deviations, but it also has a negative average drift relative to the main signal. Fig. 15 illustrates how the fusion method estimates real amount of measured variable.

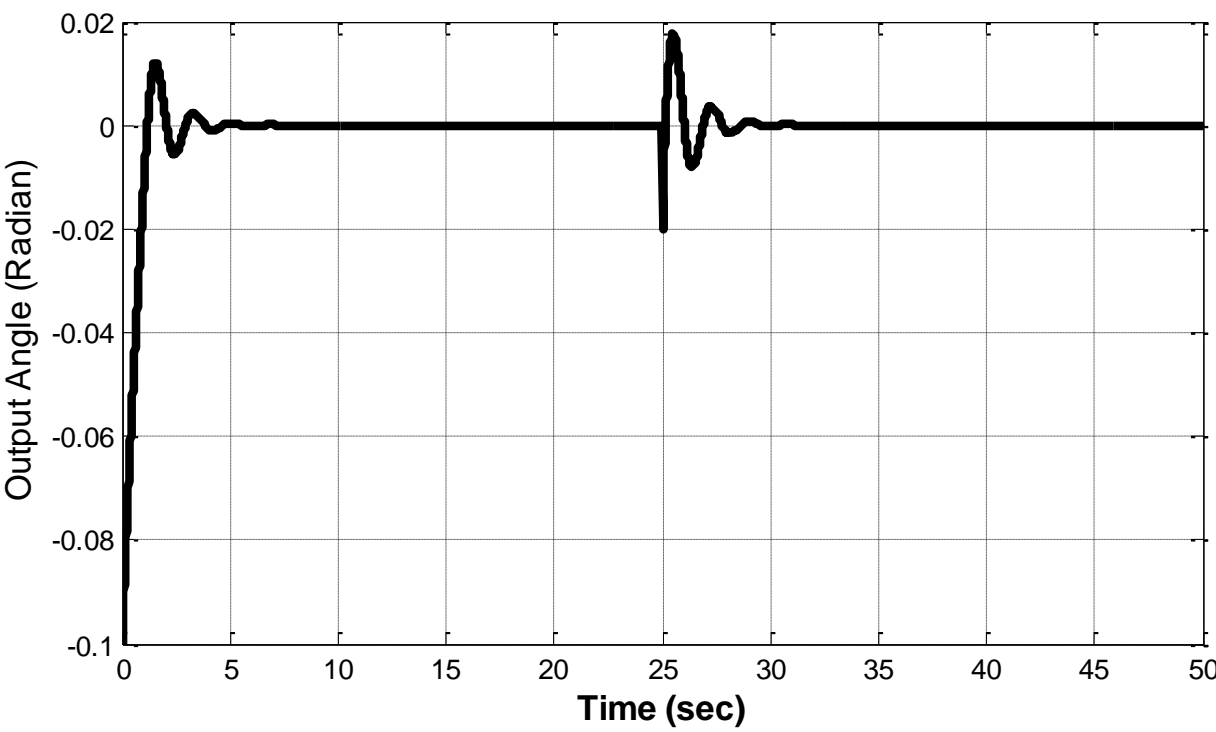

Fig. 6. System's output using ideal sensor

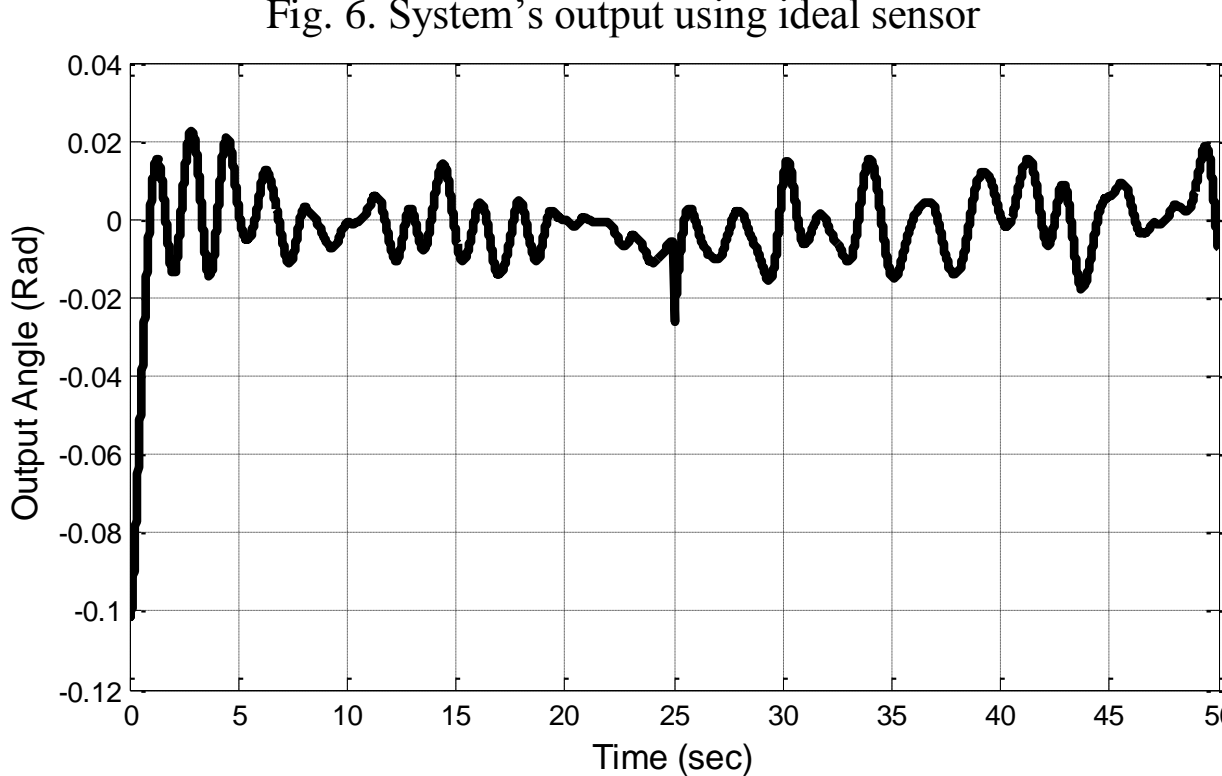

Fig. 7. System's output using sensor S1

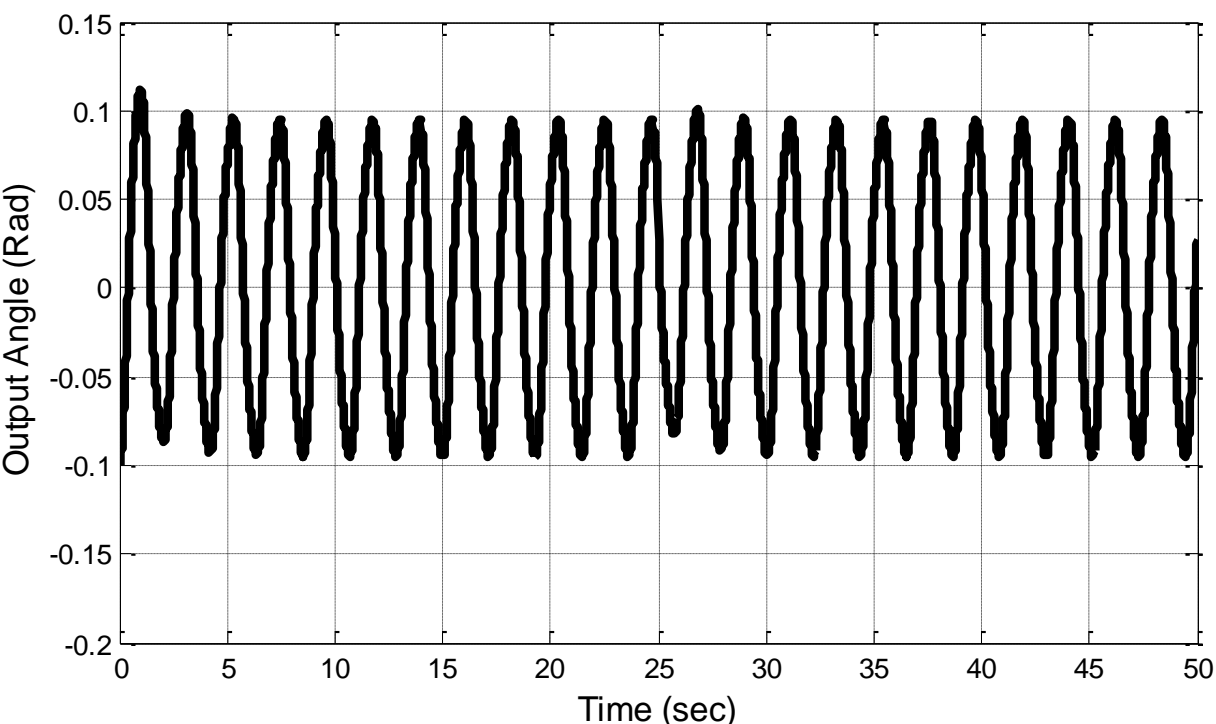

Fig. 8. System's output using sensor S2

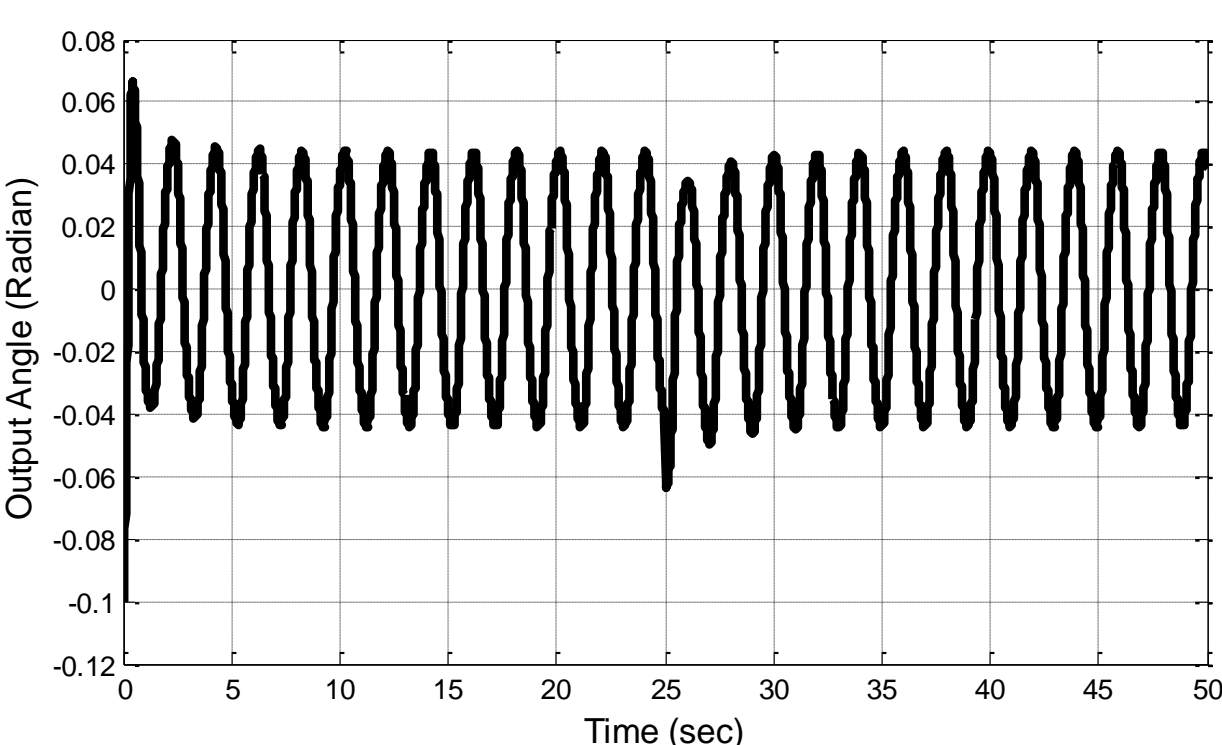

Fig. 9. System's output using ordinary average of S1 and S2 as the feedback

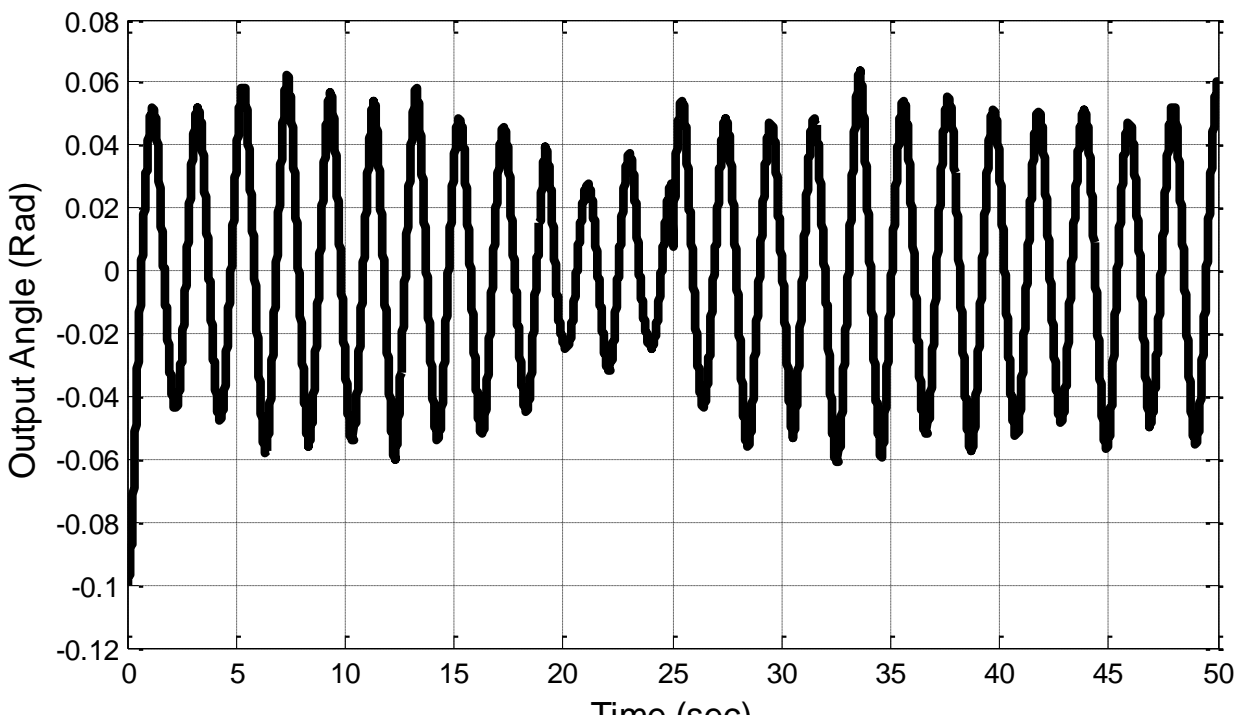

Fig. 10. System's output using fuzzy system's output as the feedback without Fuzzy Predictor and Drift output

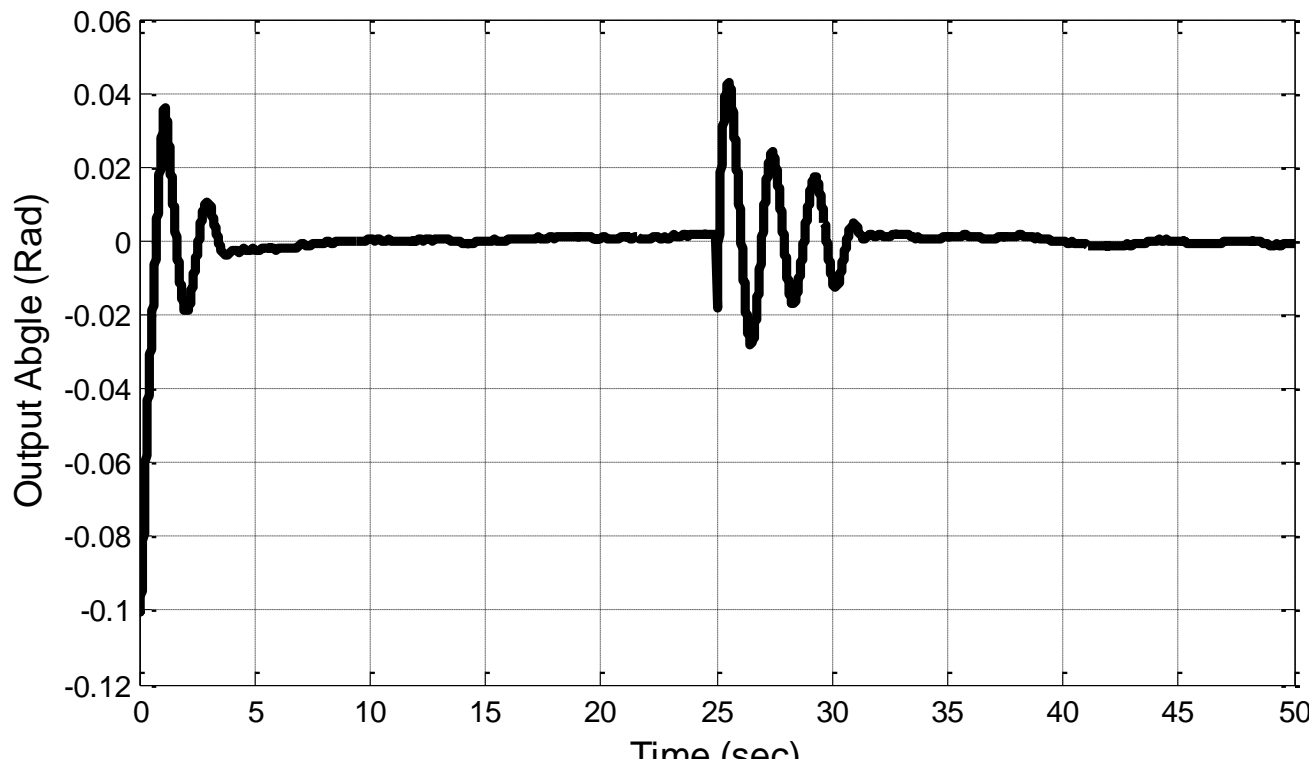

Fig. 11. System's output using fuzzy system's output without predictor

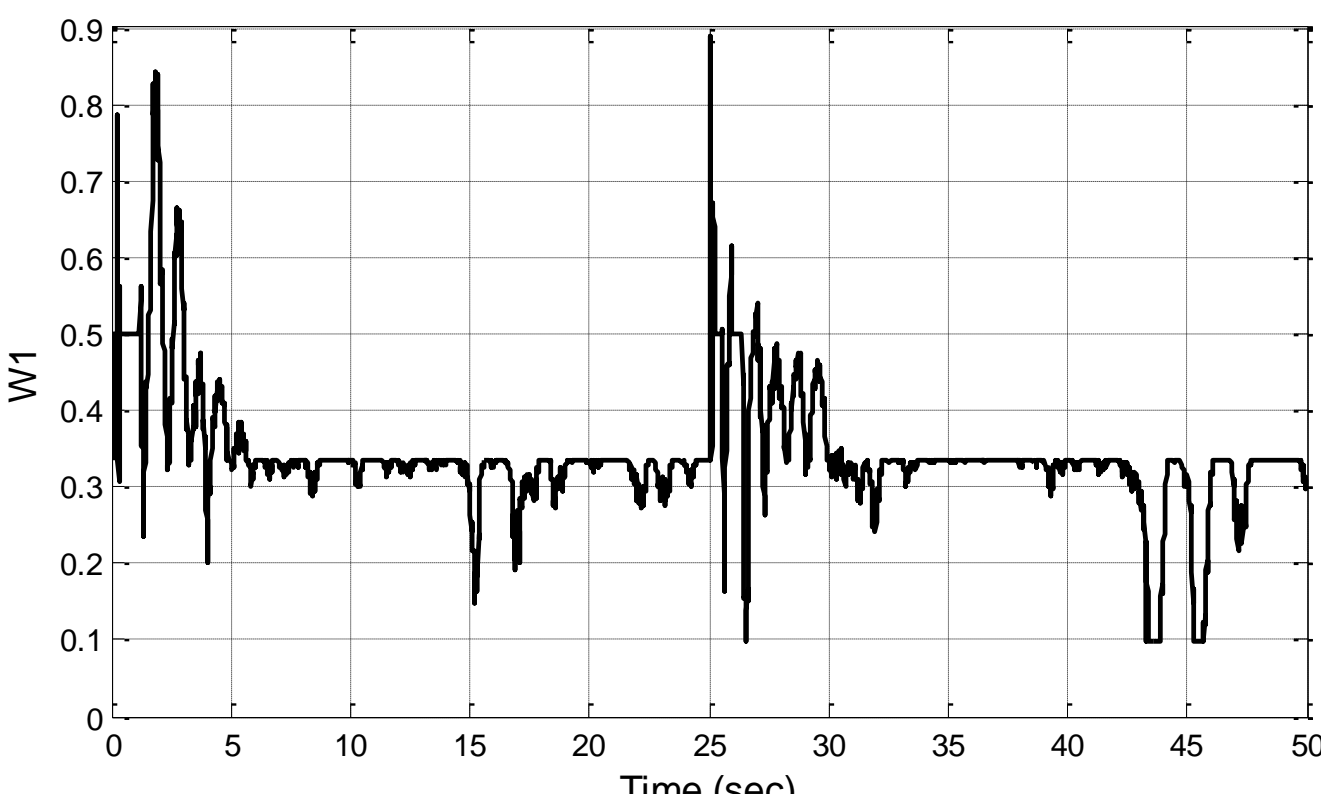

Fig. 12. W1, the weight of S1, output of fuzzy system

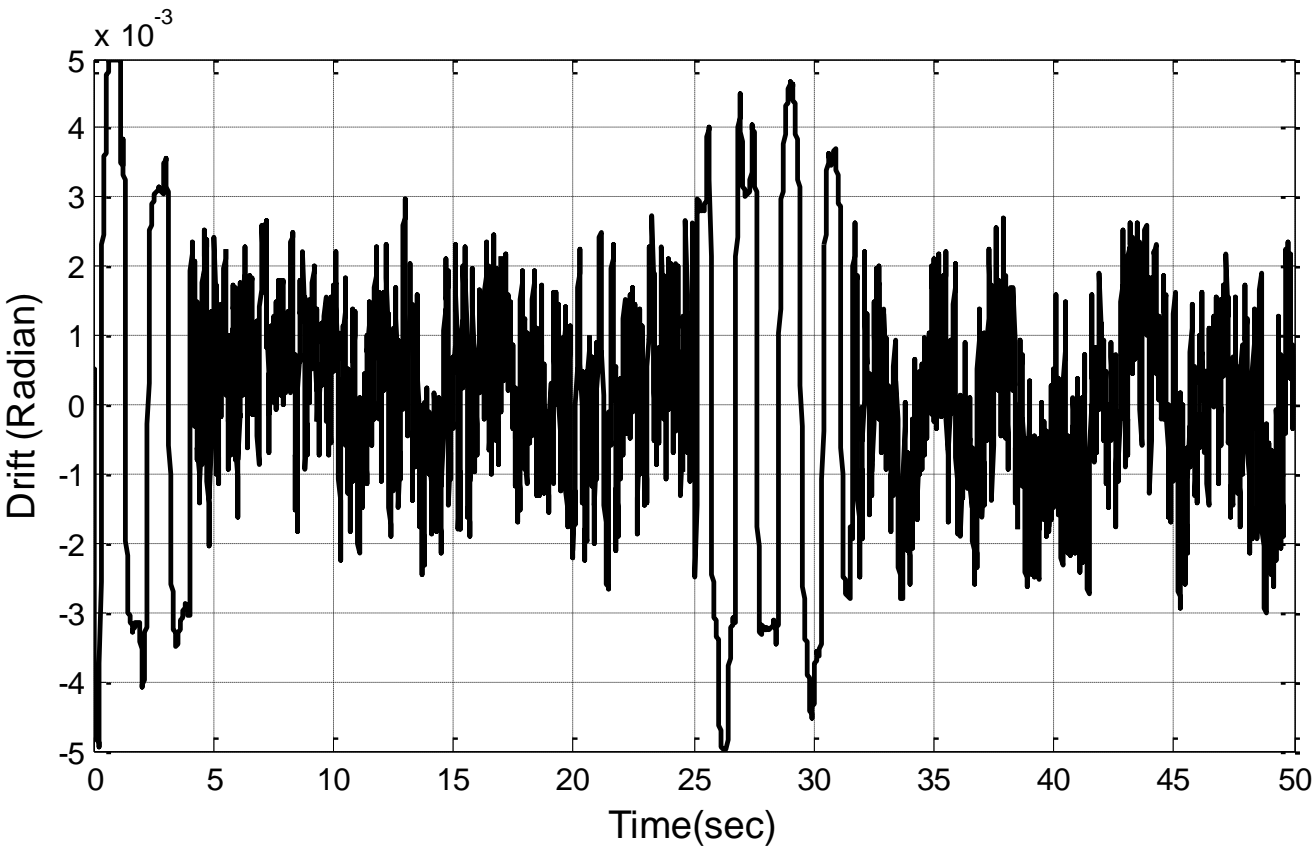

Fig. 13. "Drift" the output of rule-based fuzzy system

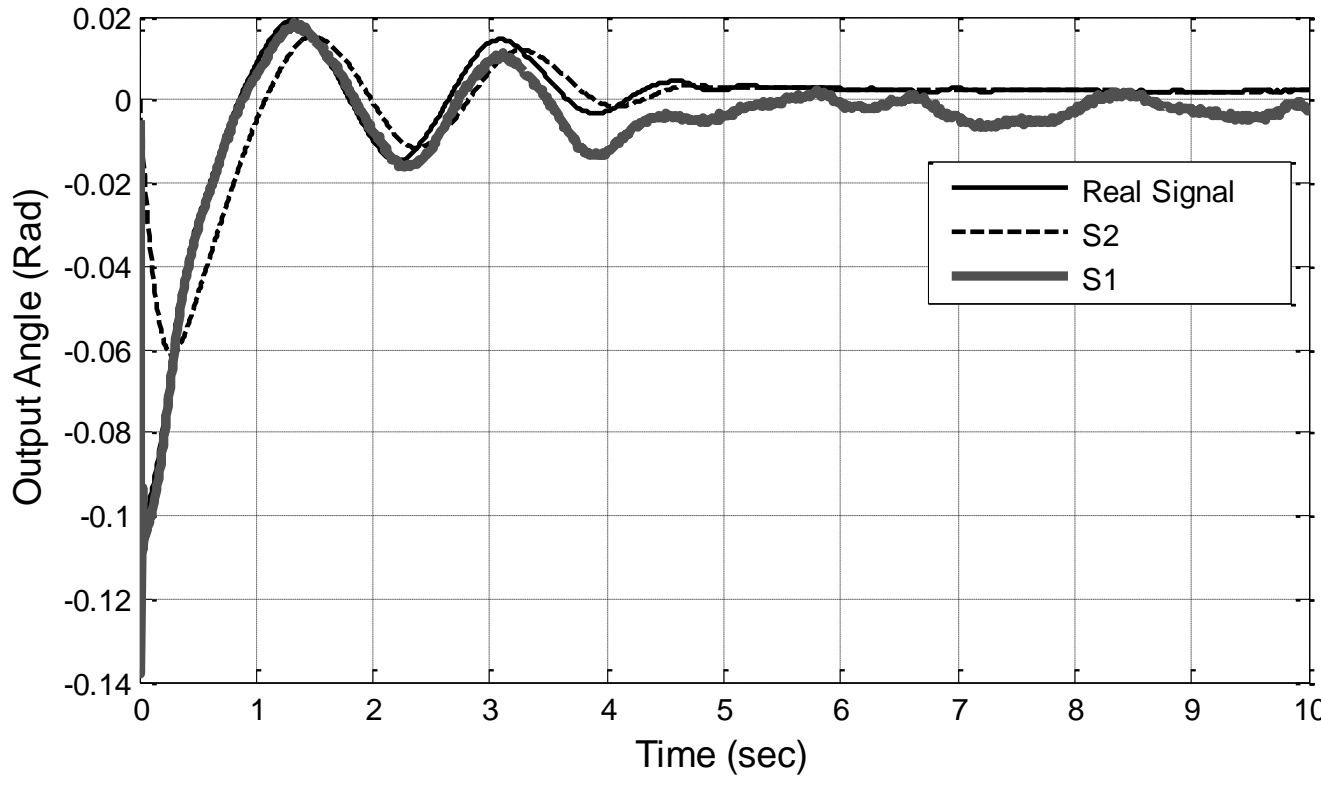

Fig. 14. Comparison between main signal and two sensors' values

In order to have a more accurate estimation and naturally a better performance of control system we can use Fuzzy Predictor as well. In our simulation, we use $n = 20$ previous samples of estimated measured. Its system's response is shown on Fig. 16. Fig. 17 shows the predicted output. A major problem with the Fuzzy Predictor is its temporal inaccurate prediction during disturbances. This problem can be detected by comparing the prediction result and the output of Fuzzy Aggregator. In this situation, the predicting system should be ignored temporarily.

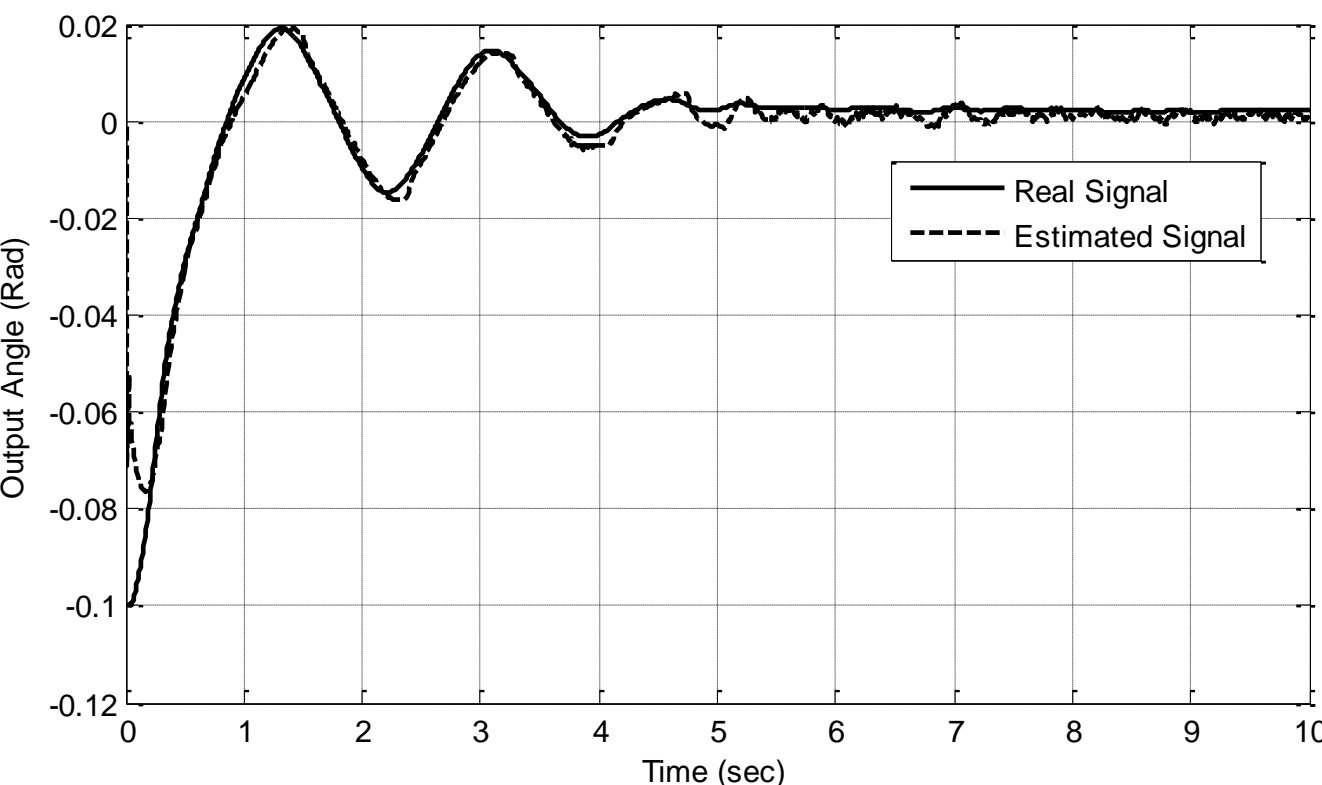

Fig. 15. Comparison between the main signal and its estimation

To evaluate system's performance efficiently, control system's response must be assessed in situations where sensors have different parameters and characteristics. In addition, we expect sensor fusion system to be robust enough to maintain acceptable performance even in cases that sensor's parameters slightly change.

To model uncertainty in sensor values, we can approximate the characteristics with a filter that has a certain bandwidth or a noise source with certain variance and arithmetic mean. But these models are not exact since sensor's characteristics can have many changes over time and hence, the system should not heavily depend on the sensors' models.

Therefore, we consider three parameters to describe sensor's behavior: Variance of deviations in sensor S1, the average of its error, and the bandwidth or time-constant of sensor S2. We use two criteria defined in (14, 15) to evaluate system's performance under alterations of these parameters.

$$IAE = \int_0^T |e(t)| dt \tag{14}$$

$$ITAE = \int_0^T t|e(t)| dt \tag{15}$$

IAE is based on absolute error of system's output over a certain period of time. ITAE is more sensitive to errors in steady state time.

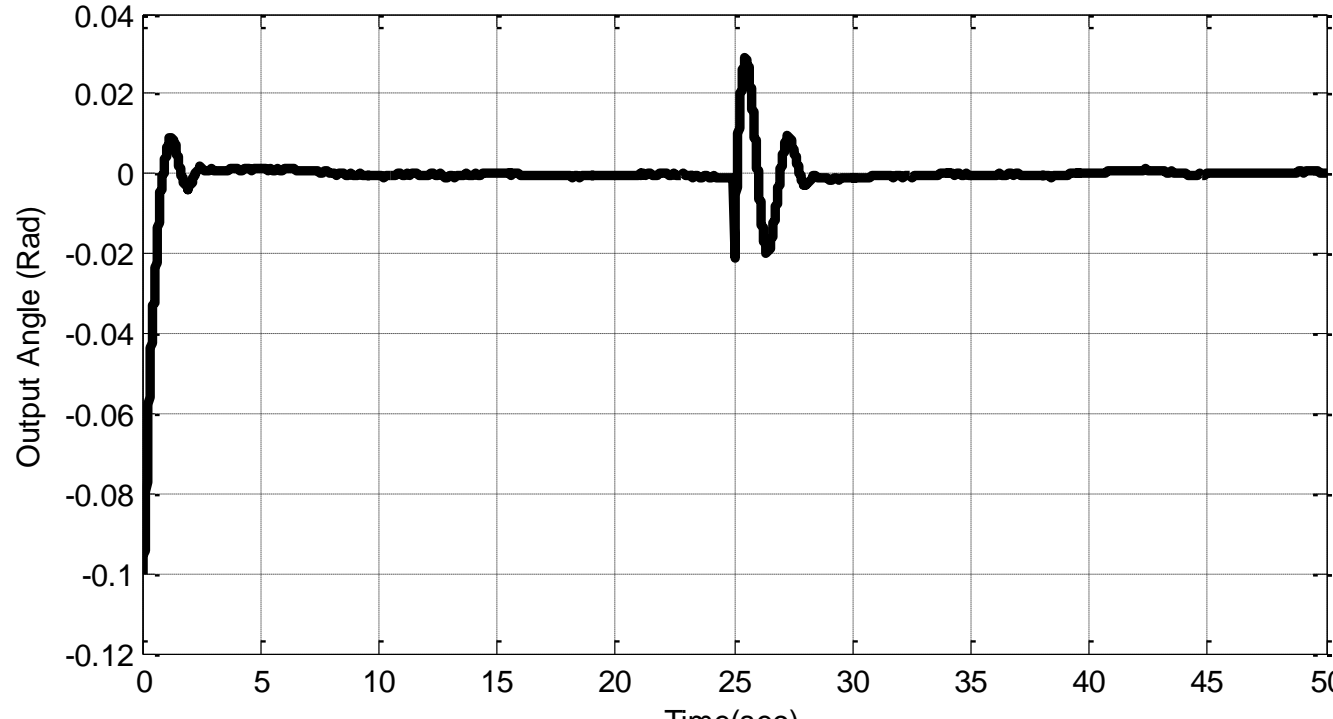

Fig. 16. The system's output using complete sensor-fusion algorithm with both fuzzy aggregator and fuzzy predictor

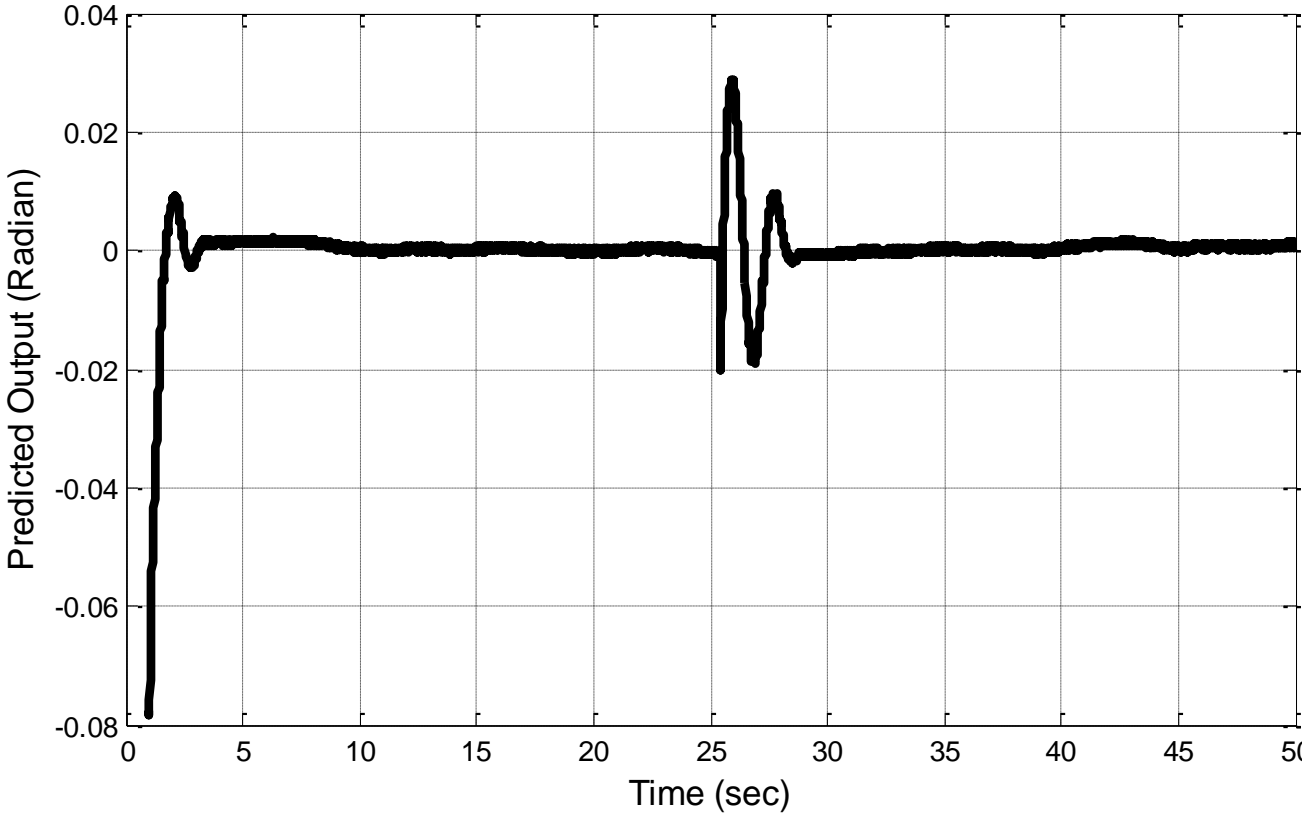

Fig. 17. Fuzzy Predictor's Output

Fig. 18 shows IAEs resulted from the control system using sensors with different bandwidths for S2 and identical S1 Sensors. The fusion algorithm and its parameters haven't been altered with changes in bandwidth S2. The results show calculated corresponding of three different sensor processing's methods: using only the sensor S2, calculating arithmetic average of S1 and S2 as feedback, and using proposed sensor-fusion method. It can be observed that in all those cases the fusion method leads to significantly better error (Smaller).

Whatever time constant the sensor S1 has had, using the single sensor or simple averaging has resulted in oscillation of the system. As shown in Fig. 19, similar results can be achieved using ITAE values. The advantages of proposed method will be more apparent when the criterion puts more values on the steady state mode of the signal. Changing second sensor's variance of deviations, the IAE and ITAE values will be identical as figures 20 and 21 illustrate. We can observe acceptable performance of the sensor-fusion algorithm even with almost wide range of error associated with S1. As shown in figures 22 and 23, the method could offer high advantages even in cases that S1 values include an unknown constant drift. In fact, these results demonstrate robustness of the fusion method and its relative independence from sensor's characteristics.

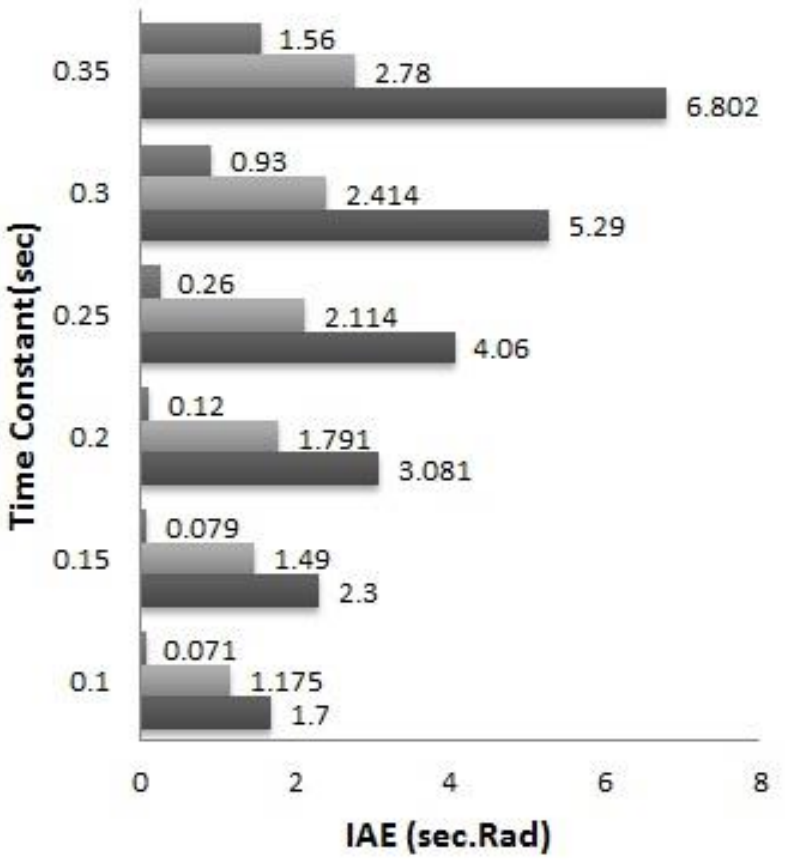

Fig. 18. IAEs obtained by sensors with different time constants for S2

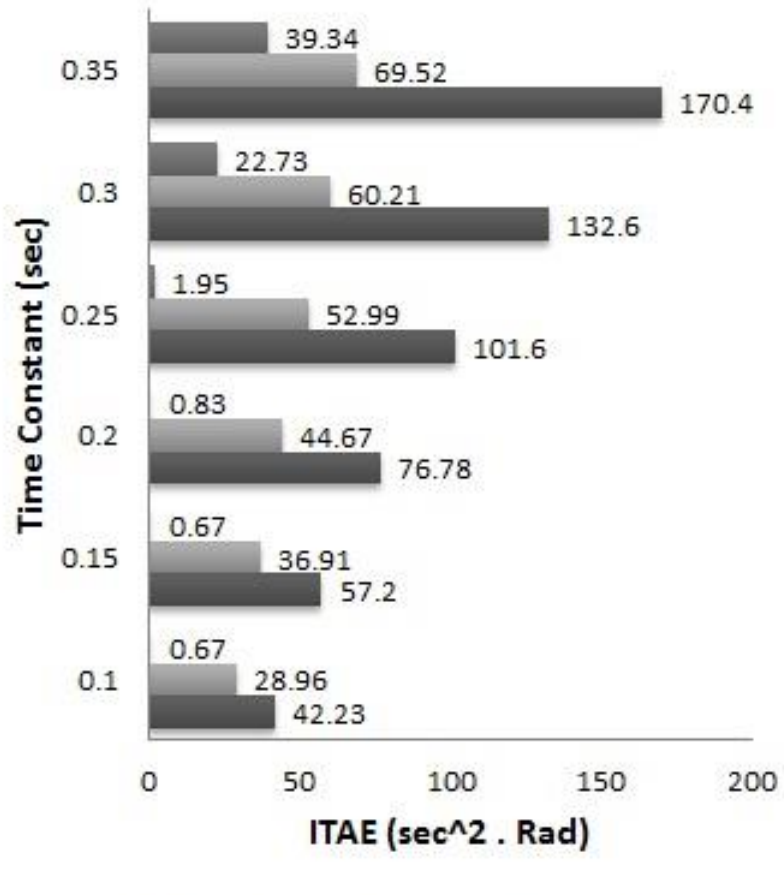

Fig. 19. ITAEs obtained by sensors with different time constants for S2

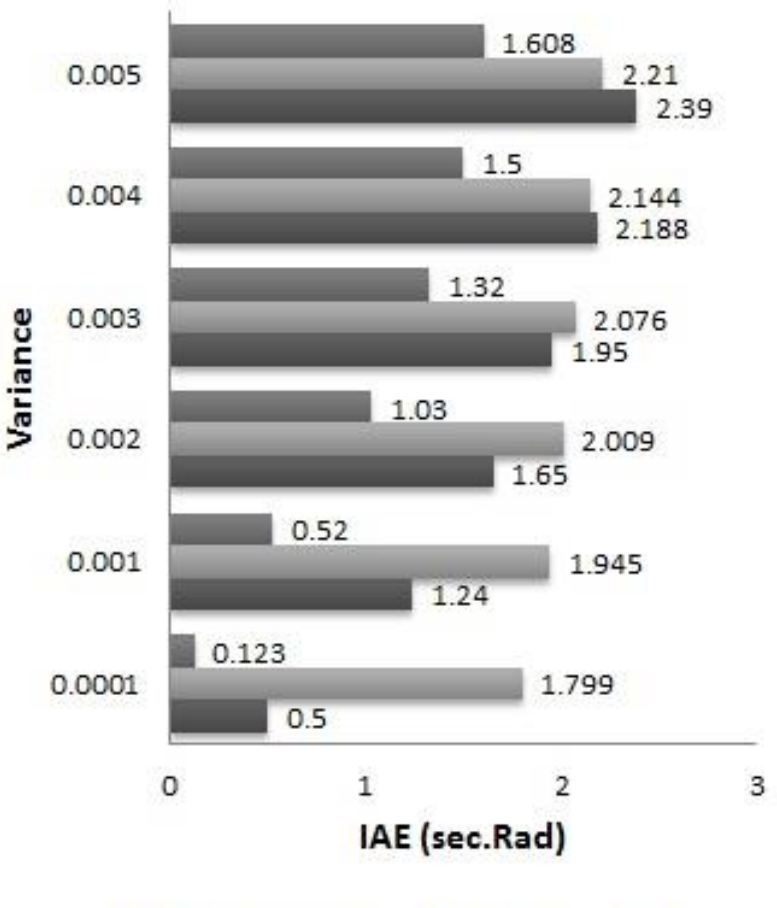

Fig. 20. IAEs obtained by sensors with different deviation's variance for S1

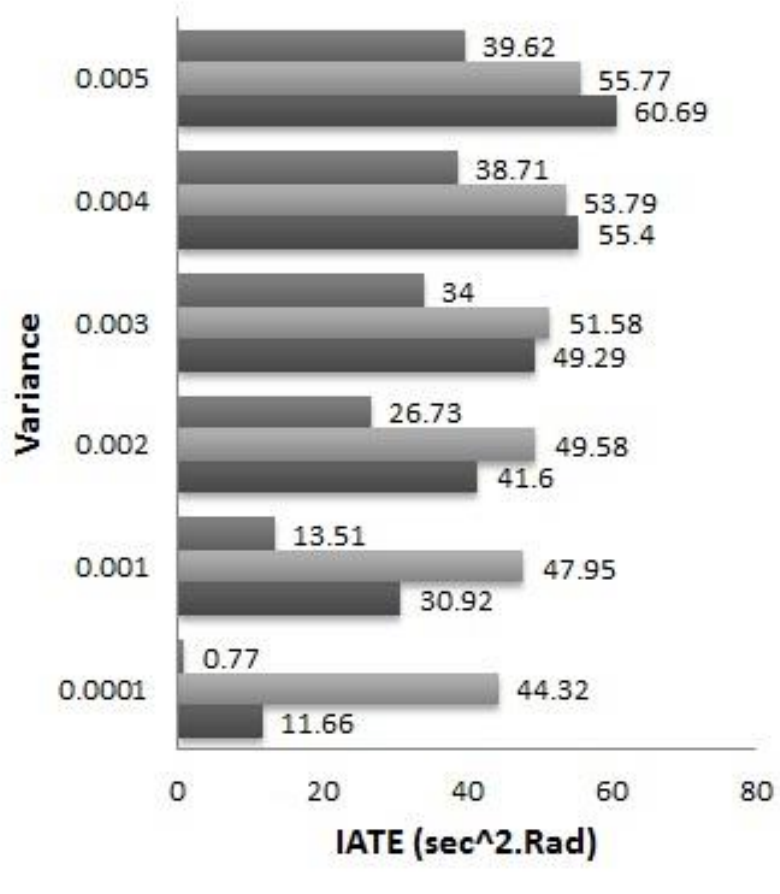

Fig. 21. ITAEs obtained by sensors with different deviation's variance for S1

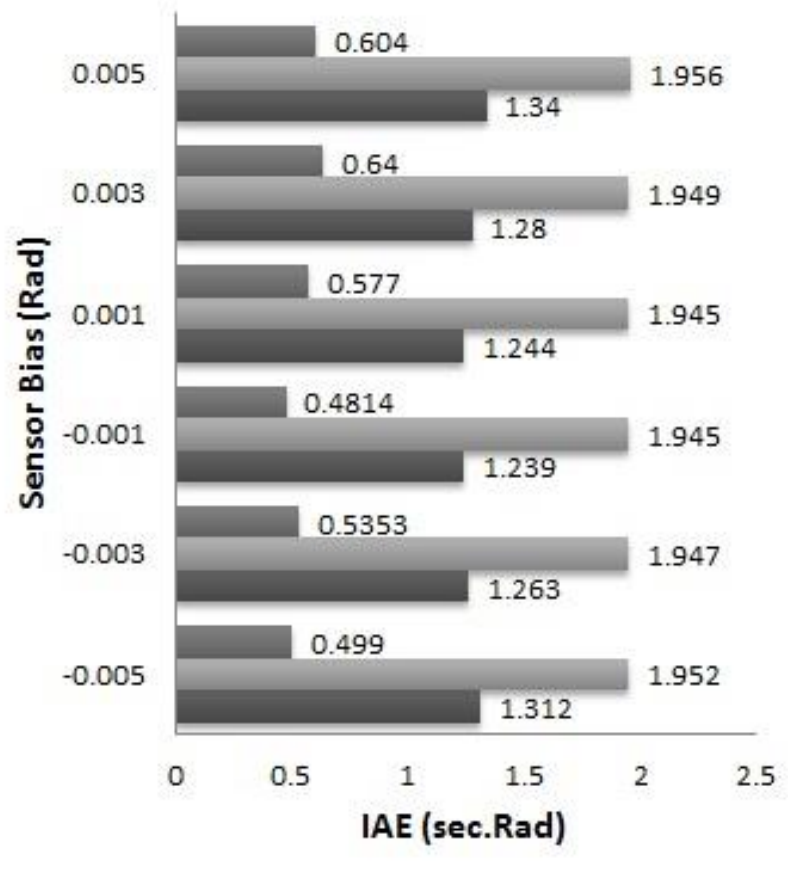

Fig. 22. IAEs obtained by sensors with different error bias for S1

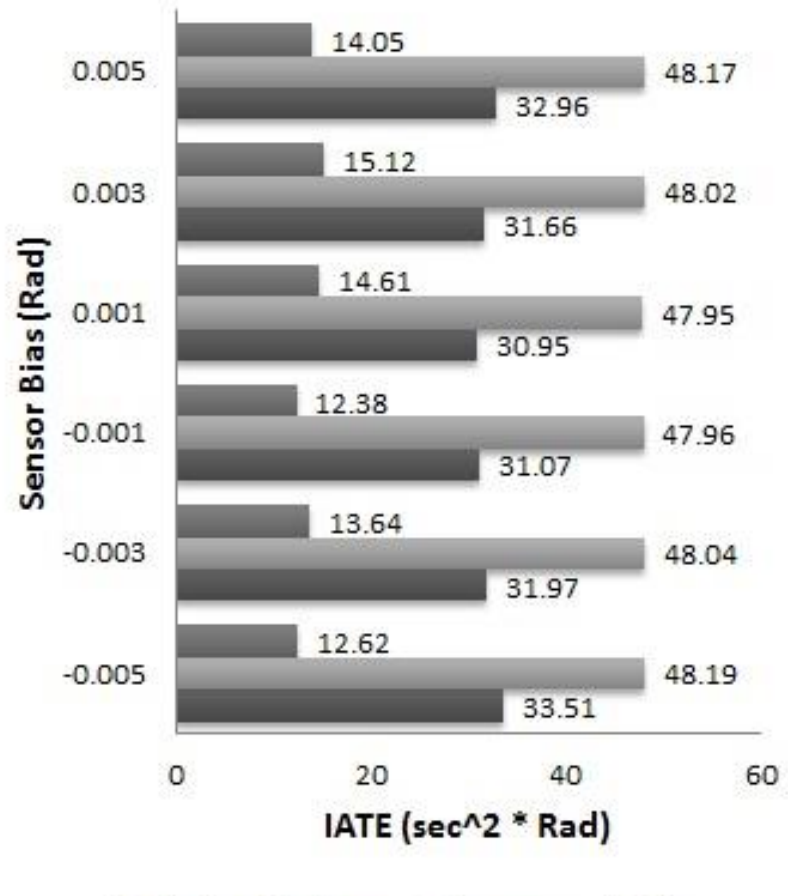

Fig. 23. ITAEs obtained by sensors with different error bias for S1

# 5. Conclusion

We presented a new approach to fuzzy sensor data fusion. Our proposed approach considers different characteristics of utilized sensors. Concentrating on accuracy and bandwidth, as the two most important and influential parameters of a sensor, we suggested a fuzzy method for two-sensor data fusion. It includes two different parts: Fuzzy Aggregator and Fuzzy Predictor. Fuzzy Aggregator takes advantages of a fuzzy system with appropriate fuzzy inputs, membership functions, and fuzzy rules. We discussed that the system without the Fuzzy Predictor can lead to an acceptable result in many applications, however, if the application requires extreme degree of certainty and accuracy, it can be added to the system. This addition comes with the high cost of complex calculations. Due to the great effects that measurement systems have in control applications, we have evaluated the performance of our method by analyzing results of a control system – as a benchmark – that utilizes our fusion method. It's worth to mention that the applications of our method are not restricted to control systems. By assessing control system's response, we have discussed great merits of the fusion method and its robust performance.

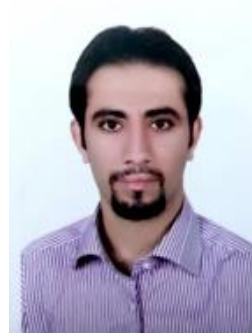

**Mohammad Amin Ahmad Akhoundi** received his B.Sc. degree in Electrical Engineering from university of Tehran, Tehran, Iran, in 2011, and his M.S degrees in Control Engineering from K. N. Toosi University of Technology, Iran in 2014.
**Email:** amin.akhoundy@gmail.com

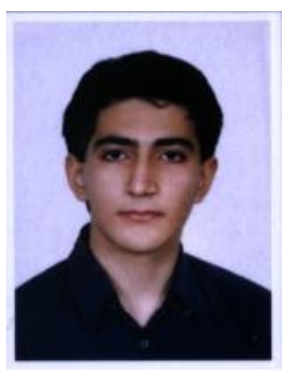

**Ehsan Valavi** received his BSc from University of Tehran in Electrical Engineering. He also holds a master’s degree in Communication Systems from Ecole Polytechnique Federale de Lausanne.
**Email:** valavi@alumni.ut.ac.ir

**Paper Handling Data:**

Submitted: 03.31.2019
Received in revised form: 08.01.2019
Accepted: 08.13.2019
Corresponding author: Ehsan Valavi
Affiliation of the corresponding author: Ecole Polytechnique Federale de Lausanne